# Manipulating carbon related spin defects in boron nitride by changing the MOCVD growth temperature

Jakub Iwański*, Jakub Kierdaszuk, Arkadiusz Ciesielski, Johannes Binder,
Aneta Drabińska, Andrzej Wysmołek

Faculty of Physics, University of Warsaw, Pasteura 5, 02-093 Warsaw, Poland
* Jakub.Iwanski@fuw.edu.pl

## Abstract

A common solution for precise magnetic field sensing is to employ spin-active defects in semiconductors, with the NV center in diamond as prominent example. However, the three-dimensional nature of diamond limits the obtainable proximity of the defect to the sample. Two-dimensional boron nitride, which can host spin-active defects, can be used to overcome those limitations. In this work, we study spin properties of $sp^2$-bonded boron nitride layers grown using Metal Organic Chemical Vapor Deposition at temperatures ranging from 700 °C to 1200 °C. With Electron Spin Resonance (ESR) we show that our layers exhibit spin properties, which we ascribe to carbon related defects. Supported by photoluminescence and Fourier-transform infrared spectroscopy, we distinguish three different regimes: (i) growth at low temperatures with no ESR signal, (ii) growth at intermediate temperatures with a strong ESR signal and a large number of spin defects, (iii) growth at high temperatures with a weaker ESR signal and a lower number of spin defects. The observed effects can be further enhanced by an additional annealing step. Our studies demonstrate wafer-scale boron nitride that intrinsically hosts spin defects without any ion or neutron irradiation, which may be employed in spin memories or magnetic field detectors.

## 1. Introduction

Defects in semiconductors associated with a non-zero electron spin are important for spintronic applications including spin memories, magnetic field detectors and quantum computation.[1] Epitaxially grown hexagonal boron nitride (hBN) is a very promising candidate for optoelectronic and spintronic applications since it can host defects possessing non-zero electron spin. [2–6] Their energetic levels are distributed over wide spectral range from near infrared to deep ultraviolet which makes these defects universal for different applications. [2–6]. On the other hand, due to its wide bandgap, impermeability for gases and low value of dielectric constant hexagonal boron nitride (hBN) is commonly used in devices [7,8], including for example sandwich-like der Waals heterostructures composed of different 2D materials.

[2,3[9] Such heterostructures, when involving spintronic materials, may open pathways for the miniaturization of electron spin-operated devices. A successful implementation of this idea strongly depends on the control over the formation of defects during the growth process. Therefore, it is crucial to understand the influence of growth conditions on the creation of spin defects in epitaxially grown hBN.

Recent studies showed that non-zero spin defects in hBN can be formed by proton or neutron irradiation.[10,11] Another strategy is oxygen dopant incorporation.[12,13] Optically active high spin states in electron-irradiated hBN were also investigated and attributed to negatively charged boron vacancies.[6] The mentioned studies were performed on bulk hBN grown by chemical synthesis or high-pressure high-temperature (HPHT) growth.[14] An alternative method for BN growth is metal-organic chemical vapor deposition (MOCVD).[15] This technique provides great possibilities to control the BN quality and properties.[16,17] The doping and the fabrication of BN heterostructures like diodes are also possible, however, further effort is necessary to optimize the properties and growth of the material. [18–21]

It appears to be natural to expect carbon incorporation in MOCVD growth, as carbon is unavoidable, being a result of the decomposition of the metal-organic precursor. Since carbon is situated in between boron and nitrogen in periodic table of elements, it can easily incorporate and mix with BN composing materials like BCN, $B_4C$ or BN:C leading to significant changes in the properties of the material.[22–25] Theoretical and experimental studies showed that the generation of spin defects occurs together with a higher imperfection of the hBN crystal lattice.[26–29] It is very likely that a part of these spin defects is related to defect complexes involving carbon.[30–33] Nevertheless, the influence of BN growth conditions on the effectiveness of carbon incorporation leading to the formation of spin-active defects has not been widely studied.

In this work, we investigate the influence of the growth temperature and sample annealing on the formation of spin-active defects in BN. We chose electron spin resonance (ESR) to study defects with non-zero spins, including carbon complexes.[34–36] Previous ESR studies of g-$C_3N_4$ and BCN material showed the presence of an intense signal assigned to delocalized electrons in the conduction band of a carbon-nitride semiconductor. ESR investigations of graphene oxide showed that ESR signal from sigma electrons localized at defects and π-electrons propagated at extended aromatic rings are characterized by narrow and wide ESR lines, respectively.[37] Their g-factors are similar, while oxidation or chemical reduction could shift the g-factor of electrons.[38]The MOCVD growth can promote the formation of different carbon defects including carbon rings in BN. Our initial studies showed that smooth, thin layers of high-quality hBN do not show any measurable ESR signal. On the other hand, thicker samples of less organized BN can show the presence of ESR-active centers. Therefore, to study the influence of the structural quality of hBN on the formation of carbon-related defects we varied the growth temperature in a series of samples in the range between 700°C and 1200°C. Moreover, we studied the influence of a post-growth annealing step on the ESR signal. Experiments performed for the samples annealed at 1200°C in nitrogen atmosphere strongly suggest a reorganization of defects in the BN crystal structure, reflected by an enhancement of the ESR signal which occurs simultaneously with a decrease of its peak-to-peak (P-P) width. The ESR studies of BN were supplemented by photoluminescence (PL) and Fourier-transform infrared spectroscopy (FTIR).[16,39,40] These techniques provide valuable information on the precursor decomposition during BN growth and the material reconstruction induced by annealing procedure, which modifies the optical properties of BN. The presented data provides an optimal temperature regarding the growth of BN containing a large concentration of carbon-related ESR-active spin defects. The combination of an appropriate growth temperature together with the high temperature annealing process opens up new pathways for the fabrication of BN containing high concentrations of spin-active defects which may be important for future spintronic applications of BN in magnetic field sensing or quantum cryptography [41–46].

## 2. Methods

Boron nitride samples were grown on 2-inch c-plane sapphire substrates using an Aixtron CCS 3x2" MOCVD system. Triethylboron and ammonia were used as boron and nitrogen sources, respectively. All growth processes were carried out using the same pressure (100 mbar) and $NH_3$:TEB flow ratio (1.25). The main difference between the studied samples is the growth temperature which varied in the range from 700 °C to 1200 °C and is encoded in the subscript of the sample name: $S_{700}$, $S_{760}$, $S_{820}$, $S_{860}$, $S_{900}$, $S_{1000}$, $S_{1200}$. The temperature was controlled during the growth process with an ARGUS optical pyrometer. The large growth temperature range renders the growth protocol used in this work substantially different from typical conditions used for the growth of high-quality boron nitride epitaxial layers, which require a much larger ammonia flow and temperatures over 1300 °C [47,48]. Two samples were studied from each growth process. One sample was annealed in nitrogen atmosphere for 15 min at 1200 °C with a temperature ramp of 30 min, and the second served as reference for the as-grown material. The annealing temperature and time were the same for all samples. Later in this work, we refer to annealed samples using the letter A in the superscript of the sample name (e.g., $S_{700}^A$).

Electron spin resonance measurements were performed using a Bruker ELEXSYS E580 spectrometer operating at the X-band equipped with a ER 4116DM resonance cavity used in $TE_{102}$ mode with a resonance frequency of 9.6 GHz. All presented spectra were recalculated for the frequency 9.5 GHz. During the measurements, the modulation amplitude was set to 0.1 mT. All the herein analyzed spectra were collected with a microwave power below the saturation of the signal. The spectrometer was equipped with a liquid helium flow cryostat (Oxford ER 4112HV), allowing for measurement of the ESR sample signal for temperatures down to 5 K. The ESR signal amplitude was normalized per unit area.

Photoluminescence measurements were carried out using Renishaw inVia Systems equipped with lasers providing different excitation wavelength (532 nm, 633 nm, 785 nm). Typically a 50x objective was used to achieve light spot on the sample below 1,5 µm diameter. Excitation of the order of few a mW did not induce heating or damaging of the investigated samples.

Fourier-transform infrared spectra were collected using a Thermo Fisher Scientific Nicolet Continuum Infrared Microscope with 32x Schwarzschild infinity corrected objective (NA 0.65). To assess the homogeneity, each sample was measured at five different positions with an aperture of 70x70 µm. The spectra did not differ and therefore, the presented FTIR spectra are averaged over all 5 measurements to achieve higher signal to noise ratio.

## 3. Results

In Figure 1 we present room temperature Fourier-transform infrared spectra of the samples. The strong reflection below 1000 $cm^{-1}$ is related to the sapphire substrate [49]. The peak at about 1370 $cm^{-1}$ is related to the $E_{1u}$ vibrational mode in $sp^2$-bonded boron nitride which is evidence of a layered crystal structure of the grown material [50–52]. The peak position shifts towards higher energies with increasing growth temperatures as presented in the inset of Fig. 1a which suggests an increasing quality of the layer. Additionally, for samples $S_{700}$ and $S_{760}$ a broad signal in the range of 1000-1300 $cm^{-1}$ is observed. It can be attributed to a significant amount of C-B bonds or amorphous BN (a-BN) [53–55]. As mentioned before, the growth processes were not optimized for high crystallinity, but for high concentrations of spin defects in BN. After annealing, the main BN peak shifts, for each sample, towards the literature value for $sp^2$-BN of 1370 $cm^{-1}$ (Fig. 1b). The intensity of another peak at 800 $cm^{-1}$ correlates with the increasing growth temperature. It is even more pronounced after annealing. However, its position is constant for each of the as-grown and annealed samples. As a consequence, we do not ascribe it to any vibrational mode in $sp^2$-BN. The sharp lines at about 2350 $cm^{-1}$ and 3500-4000 $cm^{-1}$ originate from gaseous $CO_2$ and $H_2O$ present during the measurement.[56]

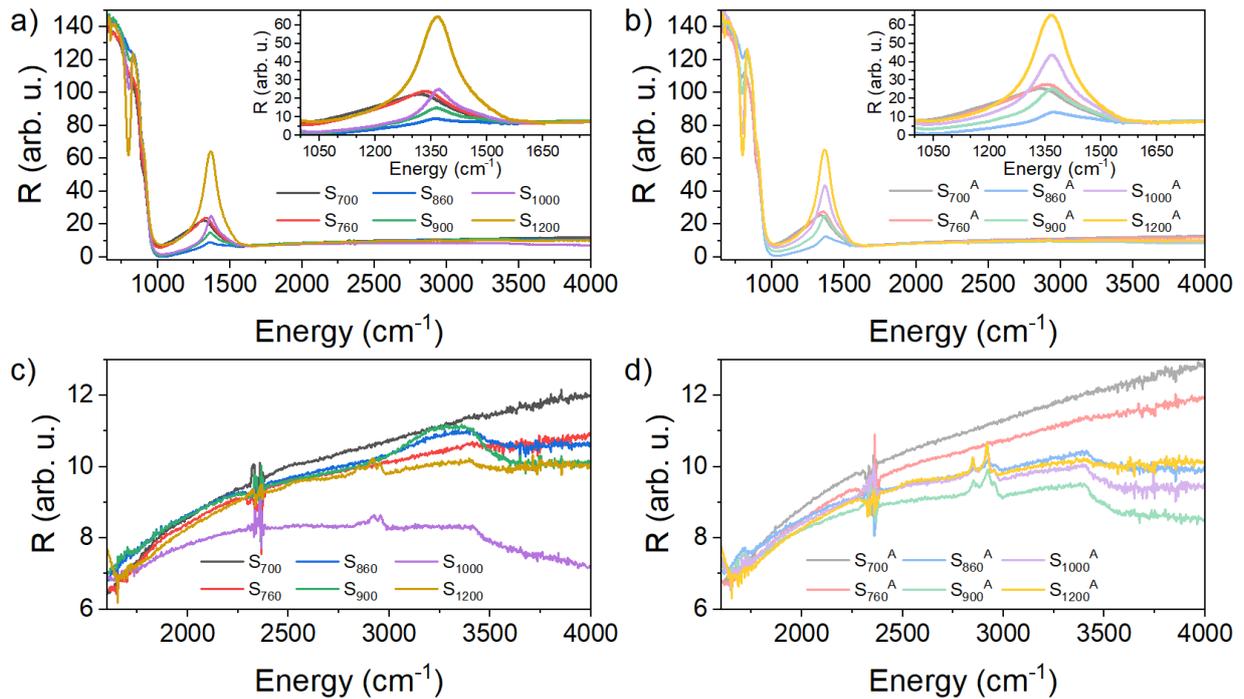

Figure 1: Room temperature Fourier-transform infrared spectra for as-grown a) and annealed b) samples. The insets present a zoom on the $E_{1u}$ vibrational mode of boron nitride. c), d) present the spectral range (1600-4000 cm$^{-1}$) for as-grown and annealed samples, respectively.

For samples grown at higher temperatures ($S_{1000}$, $S_{1200}$) an additional set of peaks was detected in the range from 2800 cm$^{-1}$ to 3000 cm$^{-1}$ (Fig. 1c). These peaks are even more pronounced after annealing and appear also for samples grown at lower temperatures ($S_{860}^A$, $S_{900}^A$) as shown in Fig. 1d. These two features in the IR spectrum at 800 cm$^{-1}$ and 2800-3000 cm$^{-1}$ can be attributed to the vibrational modes of $CH_3$ and C-H, respectively [56,57]. Moreover, for samples grown at temperatures higher than 760 °C a broad signal at 3000-3500 cm$^{-1}$ can be observed. Its intensity changes after annealing (see Supplementary Materials). This peak can be assigned to the presence of N-H bonds in our samples.

ESR spectra of BN samples measured at 5.4 K are presented in Figure 2. An ESR signal with a g-factor close to 2 was observed for as-grown samples (except $S_{700}$ and $S_{760}$) and samples annealed at 1200 °C in nitrogen. For a magnetic field ($B_0$) applied perpendicular to the sample surface, two signals were detected at g-factors of 2.0030 and 1.9683. Angular measurements identified the second signal as coming from $Cr^{3+}$ which is a common dopant in the sapphire substrate [58]. The intensity related to the signal originating from chromium varies between samples and gives no insight into the properties of BN. Therefore, we further present only results for the magnetic field $B_0$ applied in-plane of the BN layer for which the signal from $Cr^{3+}$ moves away from the region of interest.

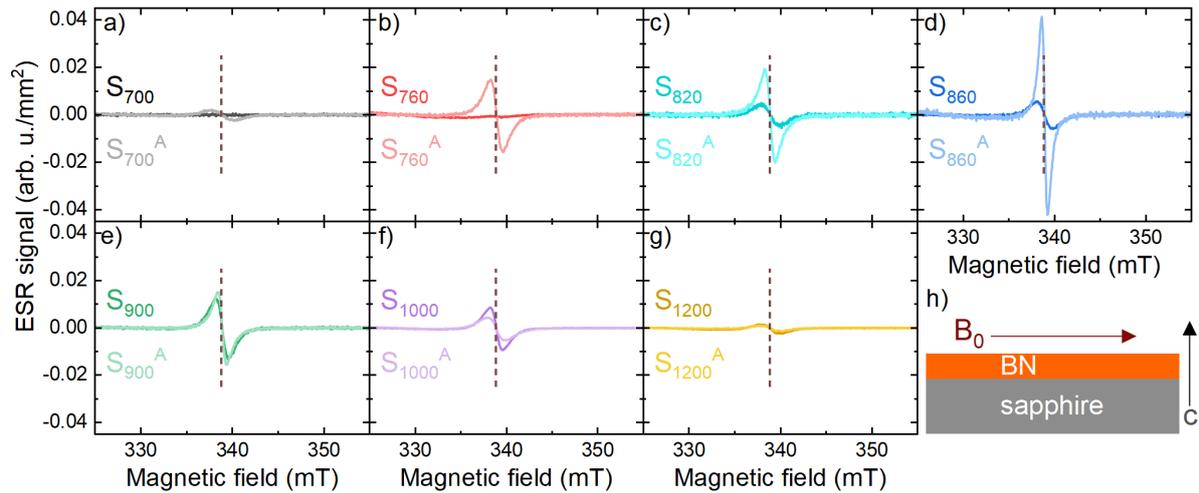

*Figure 2: Low temperature (5.4 K) ESR signal of BN on sapphire grown at different temperatures: a) 700 °C, b) 760 °C, c) 820 °C, d) 860 °C, e) 900 °C, f) 1000 °C, g) 1200 °C. Spectra with a darker color refer to as-grown samples (e.g., $S_{700}$) whereas lighter colors denote samples annealed at 1200 °C in a nitrogen atmosphere (e.g., $S_{700}^A$). The brown dashed line indices a g-factor of 2.0030. h) presents the ESR experimental setup in which the magnetic field $B_0$ was oriented in the plane of the BN layer.*

As presented in Figure 2 the peak-peak (P-P) amplitude and peak-peak (P-P) width of the ESR signal of as-grown and annealed samples strongly depend on the BN growth temperatures. The P-P amplitude is the difference between the highest and the lowest ESR signal value whereas the P-P width is the magnetic field difference between ESR peak maximum and minimum obtained from the analysis of the signal curve. The annealing affects both, the P-P amplitude and P-P width which is discussed in the next paragraph. However, the g-factor of the signal is nearly the same for all samples and is independent of the growth temperatures and annealing. It is equal to 2.0030 with a variation of 0.0004 between samples. Since pure $sp^2$-BN does not exhibit spin properties, the presence of an ESR signal and its dependence on the growth conditions suggest that some defects and impurities within the layer have to be involved. The most common defects in BN growth on sapphire by MOCVD are carbon- and oxygen-related impurities as well as boron and nitrogen vacancies [59–63]. Indeed, previous studies on g-$C_3N_4$, BCN, BNO, porous and solid BN report that such defects may lead to a ESR signal with g-factor about 2.003 [32,33,64–74]. In the following analysis we reveal the origin of the non-zero spin properties in our MOCVD BN samples as well as the impact of annealing on the ESR signal change.

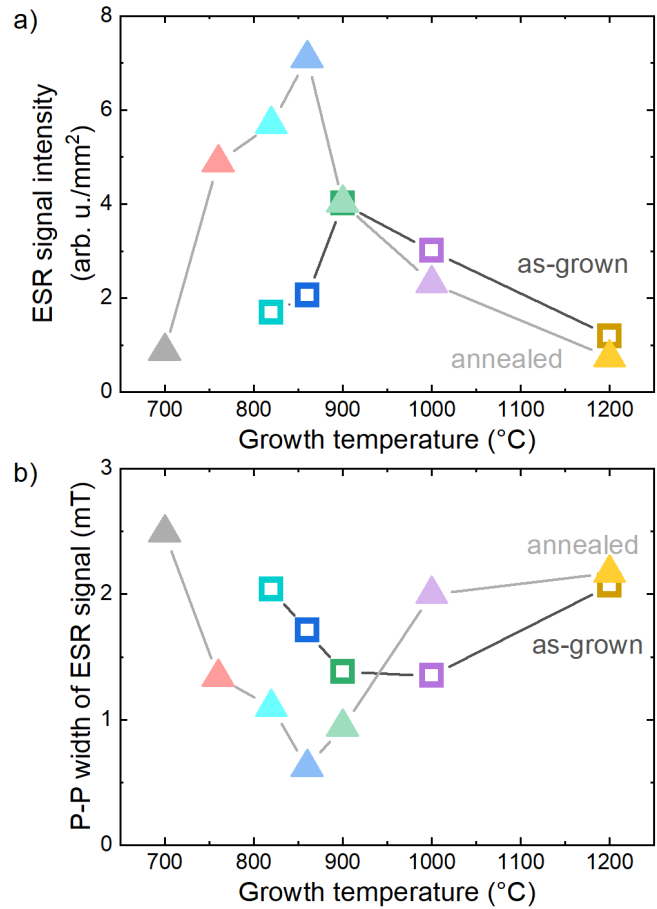

*Figure 3: Dependence on the BN growth temperatures: a) ESR signal intensity of BN, b) signal peak-peak (P-P) width. Squares and triangles denote results for as-grown and annealed samples, respectively. The color code is the same as in Fig. 2.*

At first, we investigate how the growth temperature and the annealing affect the ESR line shape. In Figure 3 we plotted two parameters: the intensity of the ESR signal and signal P-P width. For as-grown samples $S_{700}$ and $S_{760}$, no ESR signal was detected (Fig. 2a, b). Contrary to as-grown samples, an ESR signal can be detected in all annealed samples. As presented in Figure 3a, the signal appears for higher growth temperatures, its intensity increases until it reaches a maximum at a temperature of 900 °C and then it starts to decrease. Figure 3b shows an opposite trend for the P-P width. Interestingly, an enhancement of the ESR intensity is correlated with the decrease of P-P width for BN grown at 820 °C and 860 °C. The highest ESR signal intensity enhancement (up to 3.5 times) and decrease of P-P width (about 2.8 times) occurs for $S_{860}$. For sample $S_{900}$, the annealing decreases its P-P width while the ESR signal intensity is the same. For samples $S_{1000}$ and $S_{1200}$ annealing results in a decrease in intensity and an increase in the P-P width. Measurements repeated after several months showed that the effect of the thermal annealing was stable. Exposure to ambient conditions does also not influence the ESR signal significantly. This stability suggests that the observed effects are not related to gases or water adsorbed in BN. The presented results showed that the ESR signal intensity and P-P width are negatively correlated. This may be explained in terms of the influence of disorder on the ESR active center and its interaction with the BN crystal lattice which is influenced by the growth temperature. The formation efficiency of active ESR centers is highest at 860-900 °C which is evidenced by the largest ESR signal intensity for as-grown $S_{900}$ sample. Further thermal annealing in nitrogen significantly enhances the ESR signal intensity, but only in a narrow temperature range. Considering all samples, the highest ESR signal intensity was observed for the annealed sample $S_{860}^A$. Interestingly, annealing deactivates active ESR centers in BN for samples grown at temperatures higher than 900 °C.

Additional experiments as a function of microwave power indicate that the spin-lattice relaxation time is similar for most of the samples (see Supplementary Materials). However, for the sample with the highest signal intensity enhancement after annealing ($S_{860}^A$) we observed a significant reduction of the spin-lattice relaxation time. This indicates a rearrangement of the surrounding of the active ESR center during annealing. Furthermore, the ESR signal was also found to be sensitive to white light illumination. The illumination of samples $S_{860}$ and $S_{860}^A$, revealed the presence of two components in ESR signal with the same g-factor. One narrow component is sensitive to annealing and the another broad component changes upon illumination, which reveals their different nature. The two components most likely originate from different types of bonding with carbon. This aspect is discussed in detail in the Supplementary Materials.

In Figure 4 we present photoluminescence spectra obtained for samples studied with a laser excitation of 532 nm. All samples exhibit defect-related broad signals. Sharp peaks around 1.8 eV originate from $Cr^{3+}$ transitions in $Al_2O_3$ [75,76]. Spectra were multiplied by different factors for clarity to obtain the same intensity of all spectra. The total intensity differs between the samples which is due to a varying number of defects in the samples as well as the thickness of BN layers. Therefore, we will focus on general trends and the shape of the broad peaks. The intensity maximum blueshifts with growth temperature. This relation is the strongest for the 532 nm excitation. However, it was also observed for 633 nm and 785 nm excitations (see Supplementary Materials). Interestingly, the PL spectrum for each sample also blueshifts after annealing. The only exception is $S_{860}$ excited with 532 nm. The broad PL signal observed for epitaxial $sp^2$-BN in this spectral range was attributed to be related to carbon defects [59]. The observed emission is typically composed of two components with a maximum ~1.9 eV and ~2.24 eV which are ascribed to donor-acceptor pairs of defect complexes $C_B/V_BH$ and $C_B/V_B$, respectively [59]. It was also reported these defects may rearrange that during annealing. Especially, when hydrogen detaches from a boron vacancy. Furthermore, in the case of samples $S_{700}^A$ and $S_{760}^A$ an additional peaks structure is observed at an energy ~2.15 eV, which corresponds to the Raman G band, characteristic for $sp^2$-bonded graphite and graphene like structures [77,78]. This observation indicates the creation of $sp^2$-bonded carbon clusters during the annealing for these two samples. Another evidence for the rearrangement of defect complexes during annealing is the observed enhancement ($S_{820}$, $S_{860}$, $S_{900}$), weakening or no change ($S_{700}$, $S_{760}$) of the PL signal intensity. These groups of samples exhibiting different photoluminescence changes after annealing are distinguishable for all excitation energies (see Supplementary Materials).

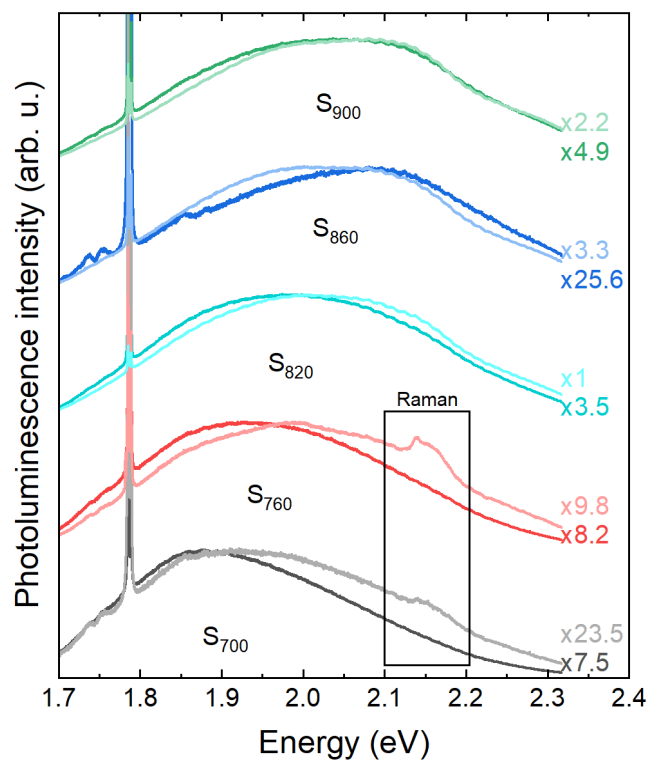

*Figure 4: Photoluminescence spectra for as-grown (darker lines) and annealed (lighter lines) exited with 532 nm laser. The spectra were shifted by a constant offset and multiplied by a scaling factor for clarity, which is shown next to each spectrum. The peak structure in the box originates from Raman G band characteristic for $sp^2$-bonded carbon clusters (see Supplementary Materials).*

## 4. Discussion

The observed ESR signal dependence on the growth temperature and the impact of annealing raises the question about its origin. The fact that pure $sp^2$-BN should not exhibit spin properties implies that a specific defect creation must be present in a narrow temperature range. The broad photoluminescence bands with maxima at around 1.9 eV, and 2.24 eV suggest that some carbon impurities have to be involved. The bands very likely originate from donor-acceptor pairs of defect complexes $C_B/V_B$-H and $C_B/V_B$, respectively [59]. We should not exclude the presence of another carbon-related defect complex $C_N/V_B$-H which is suggested by the increase of PL signal at 1.55 eV after annealing (see Supplementary Materials). Photoluminescence results in both spectral ranges suggest that carbon and hydrogen are incorporated into the crystal structure. Most likely they originate from triethylboron which is boron source during the growth process. Therefore, the pyrolysis of the used precursors cannot be disregarded. The pyrolysis of TEB starts around 550 °C and at about 700 °C boron should be released from the ethyl groups [79], which begin to degrade at this temperature. At 800 °C the ethyl groups should be fully decomposed [80]. The highest temperature is required for the pyrolysis of ammonia which starts at about 930 °C and is very effective above 1400 °C [81]. The temperature ranges mentioned are not rigid for MOCVD BN growth, because one would expect that besides the pyrolysis process other effects like chemical reactions and catalytic effects on the interface are important as well. Nevertheless, the analysis of the pyrolysis temperatures should provide an important overall trend. Considering pyrolysis temperatures of the precursors, we can extract three different temperature ranges in which we observe changes in the spin properties of the samples.

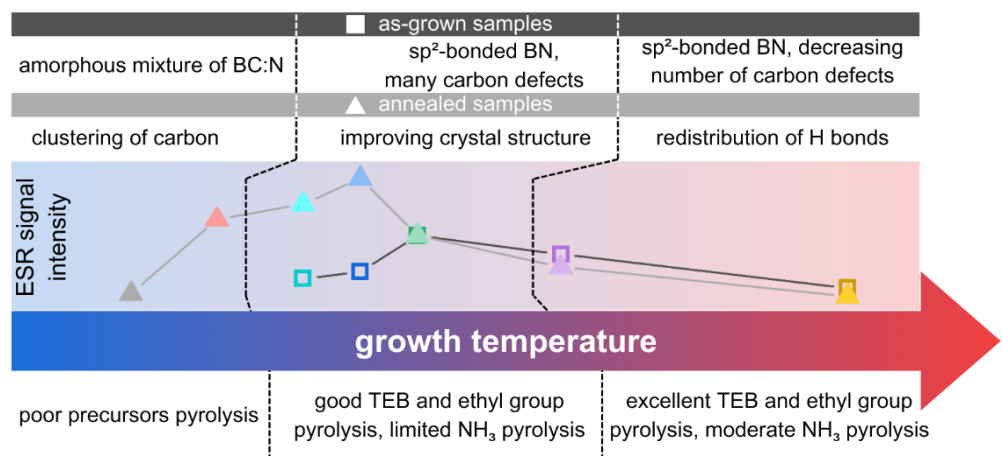

Figure 5: Graphical summary of the discussion that presents the correlation between ESR signal intensity, quality of the obtained material and pyrolysis of the precursors. Empty squares represent values for as-grown samples and filled rectangles for the samples annealed at 1200 °C.

In Figure 5 we explain the mechanism of ESR signal changes. The ESR signal intensity was taken from Figure 2 and the temperature ranges were established by the effectiveness of the precursors pyrolysis. As indicated in the literature, the ESR signal with a g-factor equal to 2.0030 may originate from delocalized electrons on carbon atoms or aromatic rings.[35,36] This interpretation is in agreement with our white light illumination experiment which revealed two components of ESR signal. A broader ESR signal component could be attributed to the electron located on the aromatic carbon rings.[37] The narrower signal component could be related to electrons located on sigma bonds between carbon atoms. The presence of such kind of defects in our samples is highly plausible, because we use triethylboron in our MOCVD growth, which consists of $sp^2$-bonded ethyl groups. Carbon with non-bonding $p_z$ orbitals can hence be incorporated into the material. At lower temperatures (T < 800 °C) precursors barely decompose. The obtained material is an amorphous mixture of BC:N ($S_{700}$, $S_{760}$). It can be noticed in FTIR results (Fig. 1) for which a peak with the maximum of ~1325 cm$^{-1}$ was observed which is interpreted as a signal from a-BN or a significant amount of C-B bonds. During annealing, carbon atoms are clustering and creating aromatic rings leading to broad ESR signal occurrence. At higher temperatures (800 °C < T < 900 °C) TEB as well as ethyl groups fully decompose. Ammonia only starts to disintegrate. Consequently, layered, $sp^2$-bonded boron nitride starts to grow which is evidenced by an FTIR peak ~1360 cm$^{-1}$. However, the material contains a significant amount of carbon. Unpaired electrons bound to carbon defects that are present within the layer lead to the appearance of the broad ESR signal. The material recrystallizes during annealing and improves its crystal structure which is evidenced by the shift of the FTIR peak to the value of 1370 cm$^{-1}$, which is very close to the value expected for hBN [50,51,82]. Correspondingly, the ESR signal enhances its intensity and reduces peak to peak width (Fig. 3). Furthermore, FTIR broad signal at 3000-3500 cm$^{-1}$ reduces and additional peaks at 2800-3000 cm$^{-1}$ and 800 cm$^{-1}$ are noticed. This draws a conclusion that hydrogen is detached from nitrogen and then bounds to carbon impurities. The proposed interpretation stems from the observation that the bond dissociation energy for H-C is higher compared to H-N [83]. It is also in agreement with the PL results since blueshift of the spectra maximum is observed which suggests hydrogen redistribution, especially on $V_BH$ defect which in fact represents N-H bonds. The attachment of hydrogen to carbon atoms concurrently results in the reduction of the number of delocalized electrons associated with carbon atoms, consequently leading to a weakened ESR signal. A further growth temperature increase allows for more effective pyrolysis of ammonia, which in consequence leads to an improvement of the BN crystal structure which is evidenced by the FTIR $E_{1u}$ peak position (Fig. 1). At the same time, less carbon built into the material, which leads to a lower ESR signal as compared to lower growth temperatures. Additional annealing leads to further decrease of the ESR signal, which can be explained by the same mechanism as before, namely a redistribution of hydrogen from bonds with nitrogen to bonds with carbon which gives a signal in FTIR at 2800-3000 cm$^{-1}$ and 800 cm$^{-1}$.

## 5. Conclusions

In this work, we studied the influence of the growth temperature on the spin properties of boron nitride. To that end, we grew a set of sp$^2$-BN by MOCVD at temperatures ranging from 700 °C to 1200 °C. After the growth samples were additionally annealed at 1200 °C in nitrogen atmosphere. We found that by using MOCVD growth, we are able to produce wafer-scale BN with non-zero spin properties without the need of a special sample treatment such as ion or neutron irradiation. ESR studies revealed that spin properties strongly depend on the growth temperature. Supported by PL and FTIR results we explain the origin of this behavior. We propose that unpaired electrons in carbon clusters within the BN layers are responsible for the spin properties of the material. By changing the growth temperature one can control the number of those defects as well as their chemical character. At low growth temperatures, inefficient precursor pyrolysis leads to low crystalline quality of the material, resulting in a significantly shortened spin relaxation time and the absence of detectable ESR signals. On the other hand, at high growth temperatures, crystallinity is improved which results in a lower number of spin-active carbon defects. Furthermore, the remaining spin centers are subsequently deactivated by hydrogen, which forms bonds with carbon, thereby removing delocalized electrons responsible for generating the ESR signal. Thus, the maximum spin related signal (observed in as-grown samples) was obtained in a narrow intermediate growth temperature range of about 860 °C. The signal amplitude can be further enhanced by annealing at 1200 °C. Concerning quantum-related spin applications our material can be easily implemented since it provides a 2-inch wafer-scale assembly of spins for growth conditions which are technologically not as demanding as in the case of high-quality boron nitride. Due to relatively low growth temperatures, our method is compliant with silicon-based technology together with a range of van der Waals heterostructures.

## Acknowledgements

This work was supported by the National Science Centre, Poland, under the decisions 2019/33/B/ST5/02766 and 2020/39/D/ST7/0281.

# Supplementary Materials

# Manipulating carbon related spin defects in boron nitride by changing the MOCVD growth temperature

Jakub Iwański*, Jakub Kierdaszuk, Arkadiusz Ciesielski, Johannes Binder,
Aneta Drabińska, Andrzej Wysmołek

Faculty of Physics, University of Warsaw, Pasteura 5, 02-093 Warsaw, Poland
* Jakub.Iwanski@fuw.edu.pl

# Variation of ESR signal as a function of microwave power and Xe lamp irradiation time

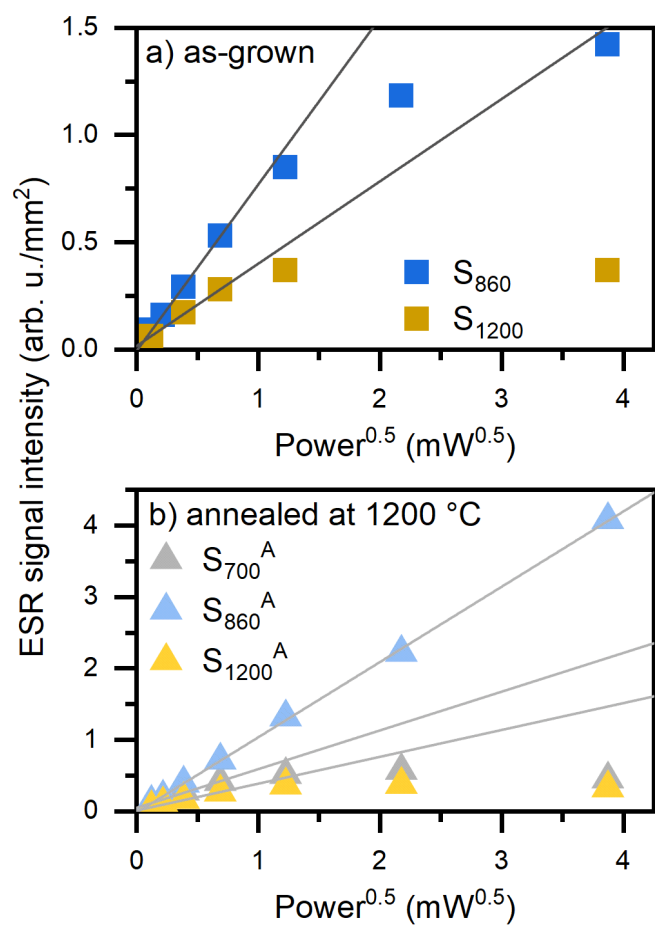

*Figure S1: Power dependence of ESR signal intensity as a function of the square of the microwave power for a) as-grown samples $S_{860}$ and $S_{1200}$, b) samples annealed at 1200 °C. Straight gray lines present fits for the linear regions of the plotted dependences.*

All discussed measurements were performed in the linear regime of the dependence between signal intensity and the square power of the microwave power. Additionally, we verified if the signal saturation threshold depends on the growth temperature and how it is affected by the annealing. For as-grown samples: $S_{860}$ and $S_{1200}$ saturation occurs for microwave power over 0.47 mW (Fig. S1a). Similar results were observed for the other as-grown samples except $S_{700}$ and $S_{760}$ where no signal was detected. After annealing, the saturation threshold for the $S_{1200}^A$ (Fig. S1b) and $S_{1000}^A$ samples was unchanged and equal to 0.47 mW. A similar value was observed for $S_{700}^A$ (Fig. S1b) as well as $S_{760}^A$. However, for samples with enhanced ESR signal intensity like $S_{860}^A$ (Fig. S1b) as well as $S_{820}^A$ and $S_{900}^A$, the saturation threshold increases to over 10 mW. Therefore, we can claim that annealing affects the spin relaxation time.

Two samples ($S_{860}$ and $S_{860}^A$) were also illuminated with white light from a xenon lamp. The samples were chosen because of the highest modification of ESR signal by annealing. The lamp power was 75 Wats. To prevent ESR cavity degradation and heating, the focus of the lamp was located a few millimeters before the optical window of the ESR cavity. This optical window has a few times lower diameter compared to the lamp beam, therefore the incident power was estimated as several mW. For higher reproducibility, the position of the lamp was not changed during both illumination experiments. Samples were illuminated while ESR signals were measured in 4-minute intervals.

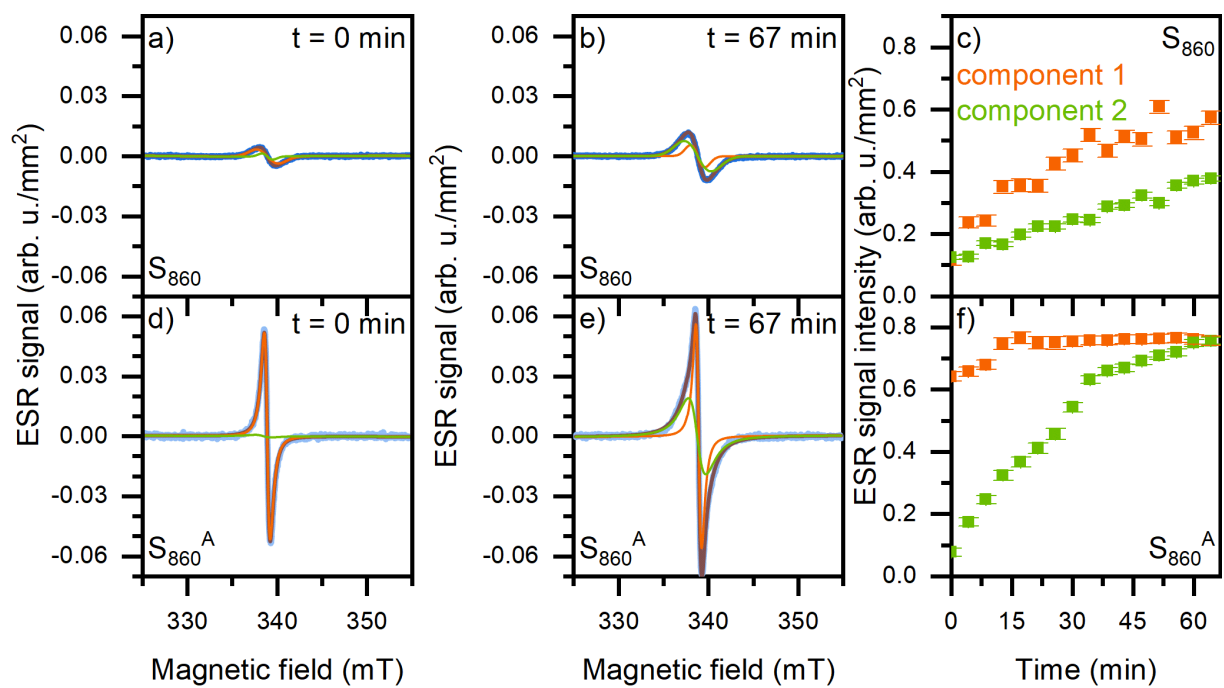

*Figure S2: Results of the experiment of $S_{860}$ and $S_{860}^A$ illumination by xenon lamp in ESR spectrometer (blue line): a) initial ESR signal of $S_{860}$ sample, orange and green lines denoted two components of the ESR signal, the brown line is the sum of components, b) ESR signal of $S_{860}$ after 4000 s of illumination, c) evolution of ESR component intensity as a function of time for $S_{860}$, d) initial ESR signal of $S_{860}^A$ sample, e) ESR signal of $S_{860}^A$ after 4000 s of illumination, f) evolution of ESR component intensity as a function of time for $S_{860}^A$ sample.*

The initial intensity of the ESR signal for $S_{860}$ is low (Fig. S2a). After more than one hour of illumination of the $S_{860}$ sample by a xenon lamp, the ESR amplitude increased 2.4 times while the signal became slightly broadened (Fig. S2b). The signal evolution of the annealed sample $S_{860}^A$ differs significantly from $S_{860}$. Initially, the ESR signal is more intense than the not annealed sample (Fig. S2d). However, the xenon lamp illumination changed the signal amplitude slightly and shape of the signal (Fig. S2e). This suggests that two components should be fitted the ESR signal of our samples. In both samples, the best fit was obtained by using the sum of Lorentz (component 1) and Gauss (component 2) derivatives with the same g-factor equal to 2.0030. The evolution of their intensities as a function of time is presented in Figures S2c and S2f for as-grown ($S_{860}$) and annealed ($S_{860}^A$) samples, respectively. In $S_{860}$ the intensity of component 1 and component 2 increased during the experiment 5 and 3 times, respectively (Fig. S2c). Their intensity ratio decreases from 0.92 to 1.52. Their peak-peak (P-P) width increases by around 1.2 times. Therefore, both component's intensities are characterized by similar responses to the illumination including increasing both intensity and P-P width. Analysis of the signal evolution in $S_{860}^A$ showed that initially component 1 dominates and the signal could also be approximated by a single Lorentz derivative curve (Fig. S2d). After 4000 s of illumination, its intensity slightly increases by a factor of 1.2 (Fig. S2f). Interestingly, the intensity of the second component increased almost 9.9 times, so the intensity of both components became comparable.

Therefore, both samples differ in response to the light illumination. In the as-grown $S_{860}$ sample, light illumination leads to an increase in the intensity of both Lorentz and Gauss signal components. It results in a moderate increase of the signal distortion. In the annealed sample $S_{860}^{A}$, the illumination affects only component 2. In the not annealed $S_{860}$ sample component 2 is more sensitive to light illumination as well. On the other hand, the most significant change in the intensity of component 1 is when comparing the signal before and after annealing. The two components of ESR signal have different nature which affects the sensitivity of each component intensity to annealing and light illumination. Therefore, component 1 can be attributed to the change of the surrounding of active ESR center meaning material structural modification, while component 2 is most likely related to the change of electronic state of the defect. Moreover, components 1 and 2 differ in width so they can be also attributed to the electrons located on sigma bonds and aromatic carbon rings, respectively.[1]

## Photoluminescence spectra for 633 nm and 785 nm excitation

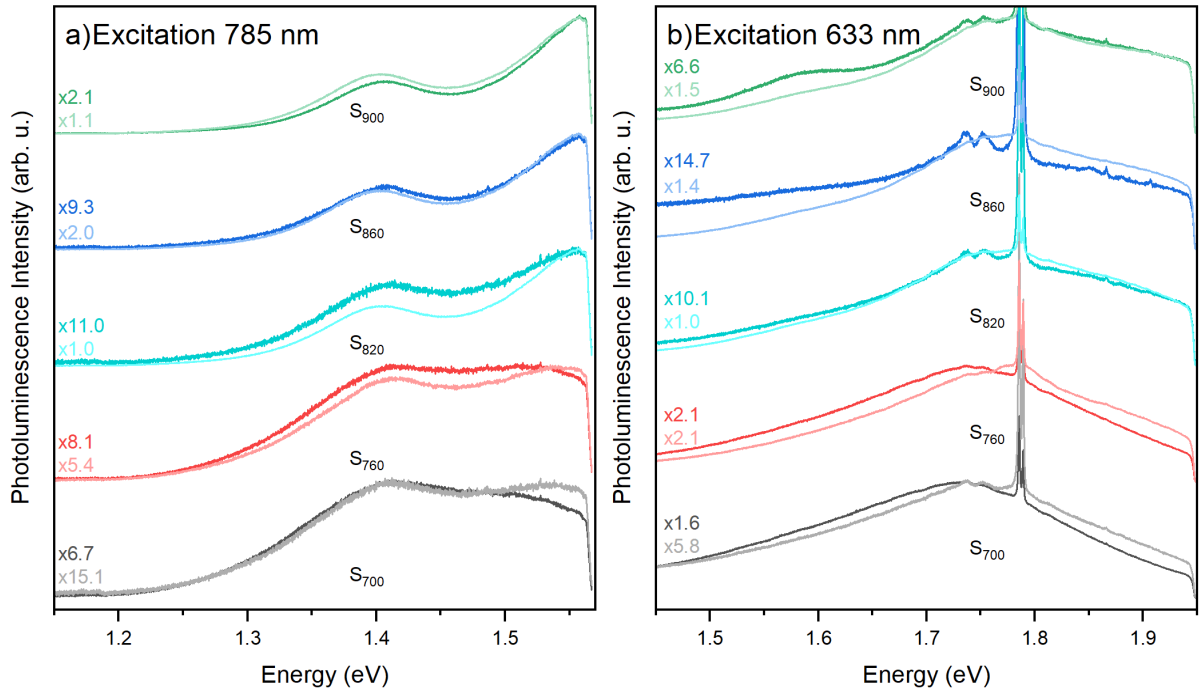

*Figure S3: Photoluminescence spectra for as-grown (darker lines) and annealed (lighter lines) samples. The spectra were shifted by constant offset and multiplied by an adequate factor for a clarity of spectrum shape change. Laser excitation energy was a) 785 nm and b) 633 nm.*

In Fig. S3 we present photoluminescence spectra for excitation energies of 785 nm and 633 nm. Intense sharp peaks at 1.78 eV with satellite peaks originate from optical transition in $Cr^{3+}$ ions in $Al_2O_3$ and their phonon replicas. The trends between spectra for as-grown and annealed samples discussed in the main text are also observed in Fig. S3. The maxima are blue shifted for annealed samples with respect to as-grown samples. Moreover, the signal intensity increases after annealing for $S_{820}^A$, $S_{860}^A$, $S_{900}^A$. It decreases for the annealed sample $S_{700}^A$. However, in contrast to the results for 532 nm excitation, the photoluminescence signal for $S_{760}^A$ stays the same for 633 nm excitation and increases for 785 nm excitation. The main peak for 785 nm around 1.55 eV can be attributed to $C_N/V_B$-H [2,3]. The increase of the population of such a defect complex that would lead to the enhancement of photoluminescence signal observed in Fig. S3a is in line with argumentation about carbon-related defects presented in the main text.

# Raman spectra for samples $S_{700}$, $S_{700}^A$, $S_{760}$, $S_{760}^A$

Raman spectra were collected using Renishaw inVia System equipped with 532 nm laser and 100x objective giving spot size of 1 μm. As presented in Fig. S4 and discussed in the main text the photoluminescence background varies for the samples. For annealed samples $S_{700}^A$ and $S_{760}^A$ additional Raman signal was observed (Fig. S4). This broad structure with a maximum of 1546 $cm^{-1}$ can be attributed to the G band characteristic for $sp^2$-bonded carbon [4,5]. The shape and broadening of the peak can be attributed to the amorphous $sp^2$-bonded carbon clusters [4]. The discussed peak is observed only for annealed samples and is more pronounced for $S_{760}^A$. This result proves that during annealing at 1200 °C in nitrogen, samples recrystallize and carbon clusters form. This process is more efficient for $S_{760}$, since it has a better crystalline quality as discussed in the main text.

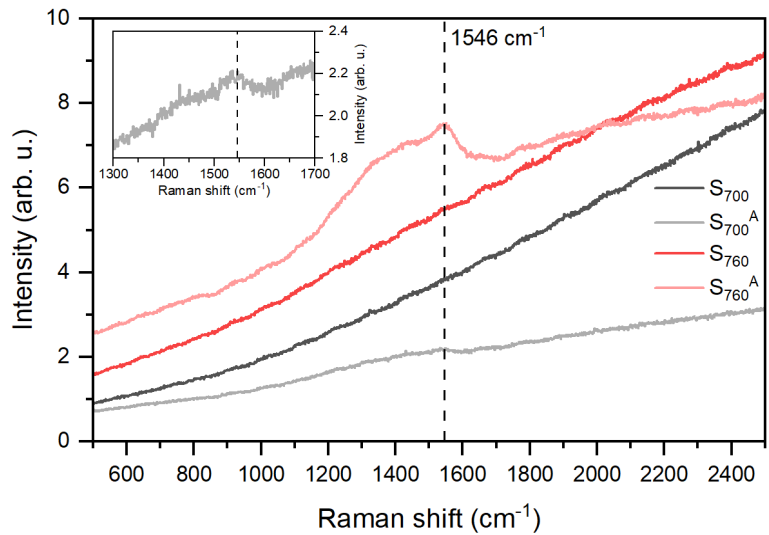

Figure S4: Raman spectra for samples $S_{700}$ (gray), $S_{700}^A$ (light gray), $S_{760}$ (red), $S_{760}^A$ (light red). Inset shows a zoom on a peak for $S_{700}^A$. Dashed reference line crosses energy of 1546 $cm^{-1}$.

# Effect of annealing on FTIR spectra (higher energy spectral range)

In Fig. S5 we present an effect of annealing on Fourier-transform infrared spectra in the range of 1750-4000 cm$^{-1}$ for the samples dicussed in the main text. The sharp lines about 2350 cm$^{-1}$ and 3500-4000 cm$^{-1}$ originate from gaseous $CO_2$ and $H_2O$ present during the measurement.[6] As discussed in the main text, the peak structure around 2800-3000 cm$^{-1}$ and broad peak 3000-3500 cm$^{-1}$ can be attributed to C-H and N-H bonds, respectively [6–9]. N-H-related signal increases its intensity with increasing growth temperature up to the maximum at 900 °C. This effect can be explained by more efficient pyrolysis of ammonia for higher temperatures which leads to incorporation of N-H/N-H$_2$ to the crystal structure. The signal decreases after annealing or at higher growth temperatures because of thermal detachment of hydrogen. At the same time, the peak structure at 2800-3000 cm$^{-1}$ appears or increases its intensity. Since this signal is related to C-H bonds, this effect can be explained by a transfer of hydrogen from nitrogen to carbon atoms, as discussed in the main text.

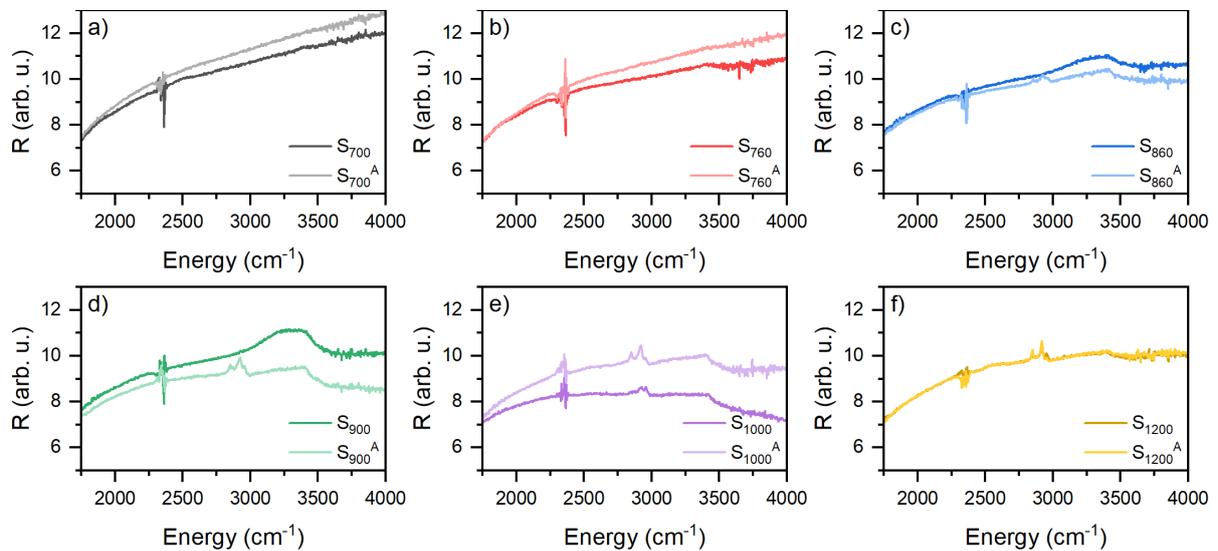

*Figure S5: Fourier-transform infrared spectra in a range of 1750-4000 cm$^{-1}$ for as-grown (darker lines) and annealed (lighter lines) samples: a) $S_{700}$, $S_{700}^A$, b) $S_{760}$, $S_{760}^A$, c) $S_{860}$, $S_{860}^A$, d) $S_{900}$, $S_{900}^A$, e) $S_{1000}$, $S_{1000}^A$, f) $S_{1200}$, $S_{1200}^A$.*